\documentclass[twocolumn,prl,amsmath,amssymb,amsfonts,superscriptaddress,floatfix,showpacs]{revtex4}

\usepackage{graphicx}
\usepackage{dcolumn}
\usepackage{bm}
\usepackage{rotating} 
\usepackage{color}

\begin{document}
\newcommand{\red}{\color{red}}

\title{
A theoretical analysis of the chemical bonding and electronic structure of graphene interacting with Group IA and Group VIIA elements
}
\date{\today}
\author{M.~Klintenberg}
\thanks{Mattias.Klintenberg@fysik.uu.se (Corresponding author)}
\affiliation{
Department of Physics and Materials Science, Uppsala University,
Box 530, SE-75121, Uppsala, Sweden
 }
\author{S.~Leb\`egue}
\affiliation{
 Laboratoire de Cristallographie, R\'esonance Magn\'etique et Mod\'elisations (CRM2, UMR CNRS 7036)
 Institut Jean Barriol, Nancy Universit\'e
 BP 239, Boulevard des Aiguillettes
 54506 Vandoeuvre-l\`es-Nancy,France
 }
\author{M.~I.~Katsnelson}
\affiliation{
 Institute for Molecules and Materials, Radboud
University Nijmegen, Heyendaalseweg 135, NL-6525 AJ, Nijmegen, The
Netherlands }
\author{O.~Eriksson}
\affiliation{
Department of Physics and Materials Science, Uppsala University,
Box 530, SE-75121, Uppsala, Sweden
 }

\begin{abstract}
We propose a new class of materials, which can be viewed as graphene derivatives involving Group IA or Group VIIA elements, forming what we refer to as graphXene. We show that in several cases large band gaps can be found to open up, whereas in other cases a semimetallic behavior is found. Formation energies indicate that under ambient conditions, sp$^3$ and mixed sp$^2$/sp$^3$ systems will form. The results presented allow us to propose that by careful tuning of the relative concentration of the adsorbed atoms, it should be possible to tune the band gap of graphXene to take any value between 0 and 6.4 eV.
\end{abstract}


\maketitle
A new derivative of graphene was recently reported\cite{elias},
where hydrogenation turned graphene, a recently discovered two-dimensional C based material,\cite{first,r1,r2,r3,klein} into what is referred to as graphane.
This compound was actually studied theoretically before the material was synthesized, and the materials phase stability as well as electronic properties (in particular the presence of a gap) were predicted from first principles theory\cite{graphane1,graphane2}.
In the GGA-based calculations of Ref.\onlinecite{graphane1,graphane2} it was found that graphane has every C atom attached to a H atom, and that the H atoms are bonded alternating on both sides of the C layer. Furthermore, it was found that graphane is a semiconductor with a rather wide energy
gap (3.5 eV). Subsequent calculations based on the GW approximation,\cite{Hedin1,Hedin2} which are supposed to result in more accurate band gaps, reported on a gap-value of 5.2 eV.\cite{graphane3}

In the experimental study\cite{elias} graphane was indeed found to have a band gap of the
electron states. Hence, it was demonstrated that the adsorption of
hydrogen turned highly conductive graphene into insulating
graphane.
The reason a band gap opens up when hydrogen is adsorbed on
graphene is that sp$^2$ bonded C atoms can achieve a higher degree of sp$^3$ bonding.
In this case three of the four covalent sp$^3$ bonds are saturated by C
atoms and the fourth covalent bond is saturated by an H atom.
The reason for the
change in conductivity is in accordance to expectations for sp$^3$
bonded carbon (e.g. diamond), which is an insulator. On an electronic structure level the transition to an insulator is connected to the fact that the p$_z$ orbitals, which in graphene form conducting states (called $\pi$ and $\pi^*$) at the Fermi level, participate in graphane in strong covalent $\sigma$ bonds, with a gap between bonding and anti-bonding states.
This finding
is illustrated in the theoretical work of
Refs.\onlinecite{graphane1,elias} and from the adsorption of
single hydrogen atoms on graphene.\cite{wessely} Moreover, in Ref. \onlinecite{misha1}, the scattering of electrons in graphene by clusters of impurities is studied and for reviews on experimental work on alkali metals on graphite, see Refs. \onlinecite{diehl,binns,yang,caragiu}.

The possibility to open up a band gap in graphene derivatives, like graphane, is very important when considering the potential of this material for electronics applications.
A natural question which then arises is if other atomic species of the Group IA elements can produce a similar behavior, potentially with a slightly smaller band gap, which is more suitable for electronics applications. It is also possible that Group VIIA atoms can adsorb on graphene, to form sp$^3$ bonds with the C atoms and with a gap in the electronic structure. For the case of single impurities, the difference in chemical bonding has been analyzed in Ref. \onlinecite{wehling}. It was found that alkali metals (except Li) form purely ionic bonds, with very small difference in total energies between different positions (on the top of a carbon atom, in the middle of the bridge between two C atoms, in the center of a C-hexagon), and the same for halogens (except F). This seems to be in agreement with the present results on complete coverage. We have here investigated the possibility of a gaped electronic structure by performing ab-initio calculations of the geometry and electr
 onic structure for the Group IA
 and Group VIIA elements adsorbed on graphene, and we will refer to this new material as graphXene (where X represents a Group IA or Group VIIA element). The theoretical calculations have been done using the generalized gradient approximation (GGA) and the GW approximation. For simplicity we have only considered the so called chair geometry\cite{graphane1} in our calculations, i.e. with one adatom attached on each C atom, alternating being above or below the C atoms (for a figure of the geometry see Fig.1 of Ref.\onlinecite{graphane1}).
We have considered several cases, for instance with Group IA elements on both sides of the C network, and with Group VIIA elements on both sides of the C network. We have also considered mixed cases where Group IA atoms are above the C network and Group VIIA atoms are below the C network.

\begin{figure}[ht]
 \begin{center}
    \includegraphics[width=0.28\textwidth, angle=0.]{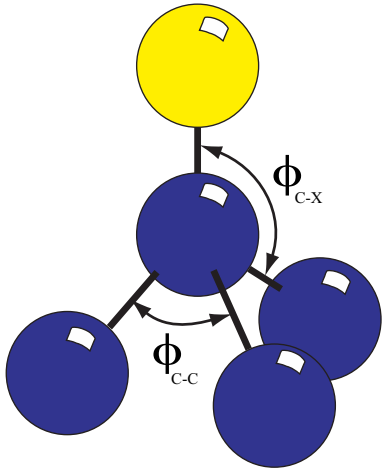}
    \caption{(Color online)
    Geometry of the C - X bonding (X is a Group IA or Group VIIA element, yellow atom, C atoms are blue). The bond-angles for the C-C and C-X bond are defined as $\phi_{C-C}$ and $\phi_{C-X}$, respectively.
 }
    \label{fig:1}
  \end{center}
\end{figure}



 We have used the VASP\cite{vasp} (Vienna Ab-initio simulation package),
 implementing the projector augmented waves (PAW) method\cite{Bloechl}, to
   compute the ground state geometry and the electronic structure.
   We obtained a reliable geometry of the structure by optimizing atomic positions, using the generalized gradoent approximation (GGA).
    For all the calculations, we
       have used a 500 eV cut-off for the wavefunction. During the optimization
       of the structures, a k-point grid\cite{Monkhorst} of $12 \times 12 \times 1$ was used.
       
Density functional theory (DFT) is a ground-state theory and can therefore not treat excited states. \cite{hohenberg1964} Equally important is that DFT does not differentiate between the two potentials V$_{N-1}$ and V$_{N}$ that should be seen by occupied- and unoccupied states, respectively. Moreover, self-interaction is not included and V$_{XC}$ does not include non-local effects nor energy- and electron density dependencies. To calculate accurate band gaps a more reliable method is needed and here we use the all-electron GW implementation based on the FP-LMTO method by Ref.\onlinecite{kotani}. The GW approximation of Hedin\cite{Hedin1,Hedin2} can be derived in a systematic way from many-body perturbation theory, and it is common to take an LDA or GGA calculation as the starting point. The result from the theoretically complex derivation is
\begin{equation}
E^{QP}_{i,k}=E^{LDA}_{i,k}+\langle i,k\mid \Sigma(E^{QP}_{i,k})-V^{LDA}_{XC}\mid i,k\rangle
\nonumber
\end{equation}
which is the quasi-particle energy resulting from first order perturbation expansion of an operator, that is made from the difference between the self-energy operator ($\Sigma=iGW$) and the LDA exchange-correlation potential ($V^{LDA}_{XC}$). $G$ and $W$ is the LDA Green's function and screened Coulomb interaction, respectively. $W$ is related to RPA polarization via the dielectric function. In this work a one-shot GW is used and we have
\begin{equation}
E^{QP}_{i,k}=E^{LDA}_{i,k}+Z_{i,k}\langle i,k\mid \Sigma(E^{LDA}_{i,k})-V^{LDA}_{XC}\mid i,k\rangle
\nonumber
\end{equation}
The quasi-particle renormalization factor $Z_{i,k}$ is given by
\begin{equation}
Z_{i,k}=\frac{1}{1-\langle i,k\mid \frac{\partial}{\partial\omega}\Sigma(E^{QP}_{i,k})\mid i,k\rangle}
\nonumber
\end{equation}
and is around 0.9 in our calculations.

There are several reviews of the GWA method in the literature \cite{review_gw1,review_gw2,review_gw3,review_gw4}. Concensus is that GWA improve the bandgap of semiconductors and insulators and the calculated GWA bandgap is often within 10\% of experiment.


In general one may expect that depending on the element which is attached to the C atom of the graphene network, a varying degree of sp$^3$ and sp$^2$ bonding will occur. The best way to characterize the degree of sp$^3$ or sp$^2$ bonding may be to use the bond angles between C-C of the graphene network as well as the bond angle between the adatom and the C atoms, i.e. the C-X angle. These angles are shown in Fig.1, where the C-C angle is labeled $\phi_{C-C}$ and the C-X angled is labeled $\phi_{C-X}$. For a purely sp$^2$ bonded system $\phi_{C-C}$ is 120 degrees and $\phi_{C-X}$ is 90 degrees. For a purely sp$^3$ bonded system $\phi_{C-C}$ is the same as $\phi_{C-X}$, with a value of $\sim$ 109.5 degrees. For intermediate cases one will observe bond angles somewhere in between these two extreme cases.
\begin{table}
\begin{tabular}{lrrrrrrrrrr}
\hline
C$_{2}$XY  & (a) & CC  & CX  & CY  & $\Phi_C$  & $\Phi_X$ & bond & Gap  & Gap & E$^F$  \\
~ & ~ & ~ & ~  & ~  & ~  & ~ & type & GGA  & GW   \\
\hline
C$_{2}$H$_{2}$  & 2.54 & 1.54 & 1.11 &      & 111 & 107 &sp$^3$ & 3.49 & 5.74 & -0.11 \\
C$_{2}$Li$_{2}$ & 2.58 & 1.53 & 2.02 &      & 116 & 77  &mix    & wm   &      & -0.64 \\
C$_{2}$Na$_{2}$ & 2.67 & 1.54 & 2.63 &      & 120 & 88  &sp$^2$ & m    &      &  0.08 \\
C$_{2}$K$_{2}$  & 2.82 & 1.65 & 2.67 &      & 118 & 81  &sp$^2$ & m    &      &  0.97 \\
C$_{2}$Rb$_{2}$ & 2.86 & 1.67 & 3.01 &      & 118 & 81  &sp$^2$ & m    &      &  1.29 \\
C$_{2}$Cs$_{2}$ & 2.94 & 1.72 & 2.92 &      & 118 & 82  &sp$^2$ & m    &      &  1.79 \\
C$_{2}$F$_{2}$  & 2.61 & 1.58 & 1.38 &      & 111 & 108 &sp$^3$ & 3.10 & 7.4  & -0.81\\
C$_{2}$Cl$_{2}$ & 2.57 & 1.47 & 3.81 &      & 121 & 90  &sp$^2$ & m    &      &  0.41 \\
C$_{2}$Br$_{2}$ & 2.69 & 1.55 & 4.18 &      & 120 & 90  &sp$^2$ & m    &      &  0.58 \\
C$_{2}$I$_{2}$  & 2.87 & 1.66 & 3.99 &      & 120 & 90  &sp$^2$ & m    &      &  1.21\\
C$_{2}$HF  & 2.57 & 1.56 & 1.10 & 1.39 & 111 & 108      &sp$^3$ & 3.11 & 6.38 & -0.47\\
C$_{2}$HCl & 2.77 & 1.66 & 1.10 & 1.76 & 112 & 107      &sp$^3$ & 0.87 & 2.91 &  0.41 \\
C$_{2}$HX\footnote{X=Cl, F alloyed 50-50} & 2.65 & 1.60 & 1.39 & 1.80 & 112 & 107  & sp$^3$ & 0.26 & ~ & -0.08 \\
C$_{2}$HBr & 2.68 & 1.57 & 1.13 & 3.70 & 118 & 98       &sp$^2$ & m    &      &  0.72\\
C$_{2}$LiF & 2.59 & 1.55 & 2.16 & 1.44 & 114 & 105      &mix    & wm   &      & -0.39 \\
C$_{2}$LiH & 2.58 & 1.53 & 2.19 & 1.14 & 115 & 104      &mix    & sm & 0.30\footnote{Using GGA gives semi-metallic behavior for C2LiH and with the GW correction a small in-direct gap opens up between K (VB) and $\Gamma$ (CB).} &  0.01\\

\hline
\end{tabular}
\caption{Geometrical and electronic structure data for all systems examined. Bond distances are given in [\AA], angles in [$^\circ$], lattice constant (a) in [\AA], gap energies in [eV], and formation energy in eV/atom. Interatomic bond lengths are denoted C-C between carbon atoms and C-X (or C-Y) for other bond distances. The systems that show a band gap are all direct gap. C$_{2}$LiH is a semi-metal (sm), C$_{2}$LiF is a borderline case sm/weak-metal and rest are metallic (m) or weak-metals (wm) at the GGA level.}
\end{table}

In Table I we list the calculated bond angles for all the systems investigated in this study. It may be seen that elemental adsorption with the same chemical species on both sides of the graphene sheet produce a situation with ideal sp$^2$ bonding for Cl, Br and I adsorption. In the cases of adsorption of Na, K, Rb and Cs one observes a chemical bonding which is almost of pure sp$^2$, whereas for Li one observes an even mixture between sp$^3$ and sp$^2$ bonding. When H or F are adsorbed, bonds which are mostly of sp$^3$ character are observed. 
Among the mixed systems, the combination of H on one side and F on the other side, or  Cl
on the other side, produce a situation with dominant sp$^3$ bonding. We will below refer to these cases as HF and HCl adsorption, respectively. A combination of Li and F, as well as Li and H, results in similar scenarios, but with a higher degree of sp$^2$ bonding. It is interesting to note that Li and H, and possibly Li and F appear to be semi-metals. Judging from Table I, we speculate that the transition in bond angle from semiconducting (or insulating) to semi-metallic is at around $114^{\circ}$ for the C-C-C bond. Finally we note that a combination of H and Br produce a situation which reflects a bonding closer to pure sp$^2$ character.

The energy of formation (E$^F$ in Table I) is calculated using E$^F$=E$_t$(graphXene)-E$_t$(graphene)-E$_t$(X)-E$_t$(Y) where E$_t$ is the total energy, graphXene=C$_2$XY where X,Y is one of H, Li, K, Na, Rb, Cs, F, Cl, Br or I. Note that E$_t$(graphene) is the total energy per unit cell, i.e. for two C atoms. It is seen from Table I that all $sp^3$ systems except C$_2$HCl have negative energies of formation, indicating that the compounds will form. The material C$_2$HCl will form if alloyed with 50\% F (lower F concentrations have not been investigated).

Moreover, if the reference state of the X or Y atoms is not the gas phase at ambient conditions, but a gas phase at elevated pressure (P) and temperature (T), a term proportional to k$_B$T$ln$P, enters the expression for the free energy, in such as way as to stabilize the phase with X and Y adsorbed on the graphene layer\cite{wassmann}.
Hence it is possible to stabilize some of the materials listed in Table I, even if $E^F > $  0.

\begin{figure}[ht]
 \begin{center}
   \includegraphics[width=0.35\textwidth, angle=-90.]{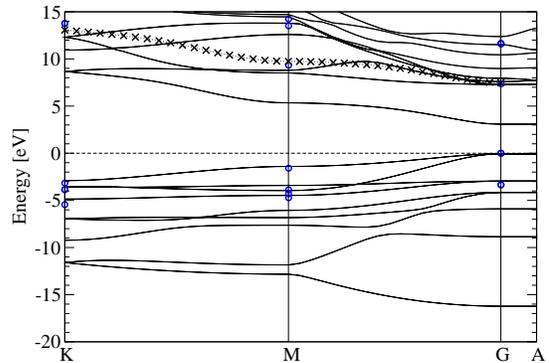}
    \caption{(Color online)
    Band structure of graphene with F on both sides of the C layer, in a chair geometry using GGA. Circles show the GW corrected energies. The rigid shift of the unoccupied states is seen when the lowest GGA conduction band is plotted but shifted to fit the GW values (dotted line with crosses).
 }
    \label{fig:2}
  \end{center}
\end{figure}

\begin{figure}[ht]
 \begin{center}
   \includegraphics[width=0.35\textwidth, angle=-90.]{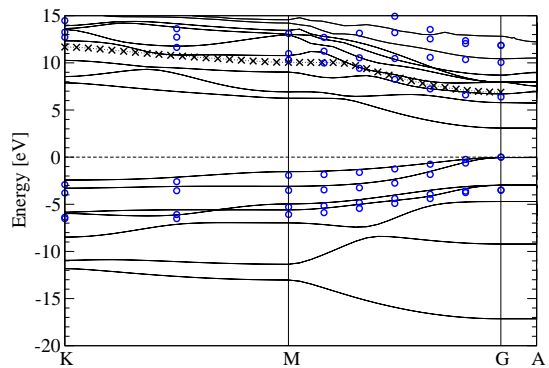}
    \caption{(Color online)
    Band structure of graphene with H on one side and F on the other of the C layer, in a chair geometry, using GGA theory. Circles show the GW corrected energies. The rigid shift of the unoccupied states is seen when the lowest GGA conduction band is plotted but shifted to fit the GW values (dotted line with crosses). }
    \label{fig:3}
  \end{center}
\end{figure}

In the cases with dominant sp$^3$ bonding it is expected that a gap opens up, whereas for the intermediate cases the situation is less clear. The electronic structures of the presently studied systems show that a band gap opens up for
H adsorption on both sides of the carbon layer as well as for F adsorption on both sides. The band structure of H adsorption has been shown before\cite{graphane1} and for this reason we do not repeat this electronic structure here. However, we show the calculated energy bands of F adsorption in Fig.2. The general shape of the energy band structure is the same for the GGA and GW calculation, except for a nearly rigid shift of the unoccupied states in the GW calculation, so as to open up a larger band gap. This nearly {\it rigid} shift is illustrated in Fig.2 (for F on both sides of the graphene layer) and Fig.3 (for F on one side and H on the other side of the graphene layer), by comparing quasi-particle GW-energies (circles) with the energies obtained by shifting rigidly the bottom of the conduction band from the GGA calculation (crosses). Both materials show a shifted GGA band structure which coinsides with the GW bands. This shift is in accordance with Fig. 1 (for diamond) of the ground-breaking paper by Hybertsen and Louie \cite{hybertsen}. An additional evidence for the nearly rigid shift in the GW calculation is illustrated in Fig. 4, which shows the quasi-particle energies plotted against the Kohn-Sham eigenvalues for C$_2$HF. From least-squares fitting (full lines) the valence bands and conduction bands are found to have the slopes 1.00 and 1.14, respectively. A slope of 1.0, with no scatter of the data around the fitted line, would indicate a perfect rigid shift for electron states of any k-point, and the data in Fig.4 gives good evidence for that the GW states are shifted more or less rigidly.

\begin{figure}[ht]
 \begin{center}
   \includegraphics[width=0.35\textwidth, angle=-90.]{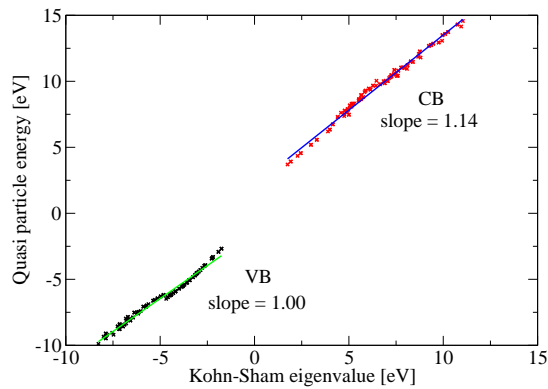}
    \caption{(Color online)
    Quasi particle energies plotted against the Kohn-Sham eigenvalues for C$_2$HF. The full lines show the least squares fit for the valence bands (VB) and conduction bands (CB), respectively. }
    \label{fig:4}
  \end{center}
\end{figure}

For the adsorption with different chemical species on the two sides of the C layer, both H-C-F and H-C-Cl are found to display band gaps. Since the two materials have electronic structures which are rather similar, we show only the GGA electronic structure of  H-C-F in Fig.3.
The H-C-F system has a direct, 3.11 eV band gap, as is seen from Fig.3. In GW theory this band gap is 6.38 eV.
The H-C-Cl case results in a smaller direct, band GGA gap of 0.87 eV, which opens up to 2.91 eV from GW theory. Therefore, a whole range of bandgaps can be induced in graphene by carefully choosing  the chemical species of the adatoms.
One can even go further by using not only two kind of dopants, but three: the alloyed structure H-C-(Cl$_{1/2}$F$_{1/2}$), gives a GGA gap of 0.26 eV, which is very different from 0.87 eV of H-C-Cl and 3.11 eV of H-C-F. This result demonstrates that the relationship between the different elements concentration and
 the bandgap is non trivial, but on the other hand, this leads to more flexibility to obtain the desired value for the band gap, a key quantity in view of using graphene in electronic devices.





In conclusion, we have demonstrated, using accurate GGA and GW calculations, that a reasonable band gap can be opened up for graphene derivatives with Group IA and Group VIIA elements, forming a new class of materials we refer to as graphXene. Depending on ad-atoms, graphXene forms sp$^2$ bonds, sp$^3$ bonds or a mixed sp$^2$/sp$^3$ binding. Several of the proposed materials have formation energies which suggest a stable configuration, where in general the sp$^3$ and mixed sp$^2$/sp$^3$ systems should be experimentally observed. We point out that although not all systems investigated here have a formation energy at ambient conditions which suggests that they should form, it is likely that elevated temperature and pressure will stabilize them.


The most important conclusion of our study is that a careful mixing of different dopants of the graphene layer enables a tuning of the band gap from 0 to $\sim$ 6.4 eV, a result which has obvious technological ramifications. For the sp$^3$ bonded, gaped systems, the C-C-C angle is close to 109.5$^{\circ}$. For the sp$^2$ bonded systems the C-C-C angle is 120$^{\circ}$, or just below this value. The intermediate cases are semi-metallic and we speculate that the semiconducting (insulating) $\to$ semi-metallic transition appears for a C-C-C angle which is around 114$^{\circ}$.

\begin{acknowledgments}
M.K. and O.E. acknowledges support from the Swedish Research Council (VR), Swedish Foundation for Strategic Research (SSF), the Swedish National Allocations Committee (SNIC/SNAC), and the G\"oran Gustafsson Stiftelse. O.E. also acknowledge support from ERC. S. L. acknowledges financial support from ANR PNANO Grant ANR-06-NANO-053-02 and ANR Grant ANR-07-BLAN-0272, and GENCI-CINES/CCRT for computer time.

\end{acknowledgments}


\end{document}